\documentclass[journal]{IEEEtran}

\usepackage{graphicx}
\usepackage{amsmath}
\usepackage{amssymb}
\usepackage{amsbsy}

\begin{document}
%
\title{The Deterministic plus Stochastic Model of the Residual Signal and its Applications}

\author{Thomas Drugman, Thierry Dutoit}%


\maketitle

\begin{abstract}
The modeling of speech production often relies on a source-filter approach. Although methods parameterizing the filter have nowadays reached a certain maturity, there is still a lot to be gained for several speech processing applications in finding an appropriate excitation model. This manuscript presents a Deterministic plus Stochastic Model (DSM) of the residual signal. The DSM consists of two contributions acting in two distinct spectral bands delimited by a maximum voiced frequency. Both components are extracted from an analysis performed on a speaker-dependent dataset of pitch-synchronous residual frames. The deterministic part models the low-frequency contents and arises from an orthonormal decomposition of these frames. As for the stochastic component, it is a high-frequency noise modulated both in time and frequency. Some interesting phonetic and computational properties of the DSM are also highlighted. The applicability of the DSM in two fields of speech processing is then studied. First, it is shown that incorporating the DSM vocoder in HMM-based speech synthesis enhances the delivered quality. The proposed approach turns out to significantly outperform the traditional pulse excitation and provides a quality equivalent to STRAIGHT. In a second application, the potential of glottal signatures derived from the proposed DSM is investigated for speaker identification purpose. Interestingly, these signatures are shown to lead to better recognition rates than other glottal-based methods.

\end{abstract}


%
\IEEEpeerreviewmaketitle


\let\thefootnote\relax\footnotetext{T. Drugman and T. Dutoit are with the TCTS Lab, University of Mons, Belgium.}

\section{Introduction}\label{sec:Intro}
\IEEEPARstart{I}{n} speech processing, the modeling of the speech signal is generally based on a source-filter approach \cite{Quatieri}. In such an approach, the source refers to the excitation signal produced by the vocal folds at the glottis, while the filtering operation refers to the action of the vocal tract cavities. In several speech processing applications, separating these two contributions is important as it could lead to their distinct characterization and modeling. The actual excitation signal is the airflow arising from the trachea and passing through the vocal folds, and is called the glottal flow \cite{Quatieri}. However, its estimation directly from the speech waveform is a typical blind separation problem since neither the glottal nor the vocal tract contributions are observable.

This makes the glottal flow estimation a complex issue \cite{Drugman-GF} and explains why it is generally avoided in usual speech processing systems. For this reason, it is generally preferred to consider, for the filter, the contribution of the spectral envelope of the speech signal, and for the source, the residual signal obtained by inverse filtering. Although not exactly motivated by a physiological interpretation, this approach has the advantage of being more practical while giving a sufficiently good approximation to the actual deconvolution problem. This paper precisely focuses on such a spectral envelope/residual signal separation and aims at finding an appropriate way of representing the residual excitation signal.

Methods parameterizing the spectral envelope such as the well-known LPC or MFCC-like features \cite{MGC}, are widely used in almost every field of speech processing. On the contrary, methods modeling the excitation signal are still not well established and there might be a lot to be gained by incorporating such a modeling in several speech processing applications.

Some efforts have been devoted in speech synthesis in order to enhance the quality and naturalness by adopting a more subtle excitation model. In the Codebook Excited Linear Predictive (CELP) approach \cite{CELP}, the residual signal is constructed from a codebook containing several typical excitation frames \cite{Maia2}. The Multi Band Excitation (MBE) modeling \cite{MBE} suggests to divide the frequency axis in several bands, and a voiced/unvoiced decision is taken for each band at each time. According to the Mixed Excitation (ME) approach \cite{ME}, the residual signal is the superposition of both a periodic and a non-periodic component. Various models derived from the ME approach have been used in HMM-based speech synthesis \cite{Yoshimura}, \cite{Maia}, \cite{Kim}. A popular technique used in parametric synthesis is the STRAIGHT vocoder \cite{STRAIGHT}. STRAIGHT excitation relies on a ME model weighting the periodic and noise components by making use of aperiodicity measurements of the speech signal \cite{STRAIGHT}. Some other techniques have incorporated into HMM-based synthesis excitation signals based on glottal flow estimates (via inverse filtering, \cite{Tuomo}) or using the Liljencrants-Fant (LF, \cite{LF}) glottal flow model (such as in \cite{Joao-LFsynth}).

In addition, excitation-based features have been shown to be useful in speaker recognition. Some of these methods aim at integrating the information of the estimated glottal flow. Based on a closed-phase linear predictive analysis, Plumpe et al. \cite{Plumpe} extracted a set of time features parameterizing the estimated glottal flow. In a similar framework, Gudnason et al. \cite{Gudnason} characterized the glottal flow by real cepstrum coefficients. Other approaches rely on the residual signal, as it is much easier to extract. In \cite{Thevenaz}, Thevenaz proposed to use the LPC coefficients of the residual signal for speaker verification purpose. More recently, Murty et al. \cite{Murty} highlighted the complementarity of the residual phase with conventional MFCCs in speaker recognition.

The goal of this article is to propose a Deterministic plus Stochastic Model (DSM) of the residual signal and to show its usefulness in both speech synthesis and speaker recognition. The potential of this model has been already investigated in \cite{DSM} and \cite{Drugman-Reco}. The main contributions of the present article are to provide a more detailed theoretical framework of the proposed DSM, to study its properties and to extend our experiments both in speech synthesis and speaker identification.

This article is structured as follows. Section \ref{sec:DSM} details the formalism of the proposed DSM of the residual signal and some of its properties are discussed. The two following sections are devoted to the applicability of this model in two fields of speech processing. First, Section \ref{sec:Synthesis} focuses on the improvement brought by the incorporation of the DSM vocoder in HMM-based speech synthesis. Secondly, two glottal signatures derived from the proposed DSM are shown in Section \ref{sec:SpeakerReco} to be useful for speaker identification. Finally, Section \ref{sec:conclu} concludes the contributions of the article.




\section{A Deterministic plus Stochastic Model of Pitch-Synchronous Residual Frames}\label{sec:DSM}

\begin{figure}[!ht]
  \centering
  \includegraphics[width=0.45\textwidth]{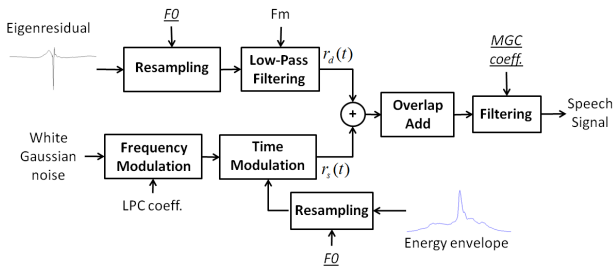}
  \caption{Workflow of the DSM vocoder. Input features (indicated in italic and underlined) are the target pitch $F_0$ and the MGC filter coefficients. All other data is precomputed on a training dataset. A thorough description of the DSM vocoder is given in Section \ref{ssec:vocoder}.}
  \label{fig:Vocoder}    
\end{figure}

The vocoder based on the proposed Deterministic plus Stochastic Model (DSM) of the residual signal is presented in Figure \ref{fig:Vocoder}. DSM stems from an analysis performed on a speaker-dependent set of residual frames that are synchronous with a Glottal Closure Instant (GCI) and whose length is set to two pitch periods (see Section \ref{ssec:residual}). This process is required for matching residual frames so that they are suited for a common modeling. Each residual frame $r(t)$ is modeled as the sum of two components: \emph{i)} a low-frequency deterministic component $r_d(t)$, based on a waveform obtained via PCA decomposition and detailed in Section \ref{ssec:deterministic}, and \emph{ii)} a high-frequency noise component $r_s(t)$ modulated both in the time and frequency domains and described in Section \ref{ssec:stochastic}. These two components are separated in the spectral domain by a particular frequency called \emph{maximum voiced frequency} and noted $F_m$, as explained in Section \ref{ssec:Fm}. The deterministic and stochastic components are then added, and the resulting GCI-synchronous residual frames are overlap-added. The reconstructed residual signal is finally the input of the filter (modeled in the case of Figure \ref{fig:Vocoder} via the Mel-Generalized Cepstral (MGC, \cite{MGC}) coefficients) to give the synthesized speech signal. Note that a more thorough description of the DSM vocoder will be given in Section \ref{ssec:vocoder}. Finally, two important properties of the DSM, namely speed of convergence and phonetic independence, are respectively discussed in Sections \ref{ssec:convergence} and \ref{ssec:independence}.

The idea of considering two separate subbands in the spectral domain delimited by the maximum voiced frequency $F_m$ has been proposed in the Harmonic plus Noise Model (HNM, \cite{HNM}). Originally, HNM models speech (and not the residual signal) via two components: a harmonic model (sum of sinusoids at multiples of the fundamental frequency) is used for the low-frequency contents, and the contribution beyond $F_m$ is considered as noisy. In \cite{Vincent}, Vincent et al. suggested for voice transformation an Auto-Regressive with eXogenous Liljencrants-Fant input (ARX-LF), where the LF residue is modeled by a HNM. The LF residue was further modeled in \cite{Rosec-ICASSP09} respectively by a modulated noise and by a harmonic model, showing an advantage for the latter approach in terms of quality in an analysis-synthesis context. The proposed DSM mainly differs from the previous models in the following points: \emph{i)} DSM models the residual signal obtained after removing the contribution of the spectral envelope, \emph{ii)} DSM analysis and synthesis is performed GCI-synchronously for both deterministic and stochastic components, \emph{iii)} the deterministic part consists of an orthonormal decomposition of the residual frames, \emph{iv)} as explained in Section \ref{ssec:vocoder}, the only excitation parameter of DSM is the fundamental frequency, all other data being precomputed for a given speaker with a given voice quality.

\subsection{A Dataset of Pitch-Synchronous Residual Frames}\label{ssec:residual}

\begin{figure}[!ht]
  \centering
  \includegraphics[width=0.45\textwidth]{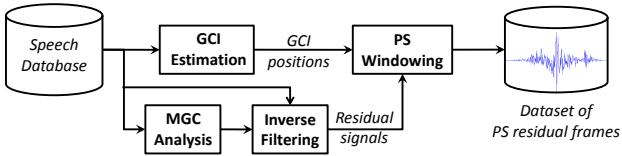}
  \caption{Workflow for obtaining the pitch-synchronous residual frames.}
  \label{fig:PSframes}
\end{figure}

The workflow for obtaining pitch-synchronous residual frames is presented in Figure \ref{fig:PSframes}. For this, a speaker-dependent speech database is analyzed. First the locations of the Glottal Closure Instants (GCIs) are estimated from the speech waveform using the SEDREAMS algorithm \cite{SEDREAMS}. GCIs refer to the instants of significant excitation of the vocal tract. These particular time events correspond to the moments of high energy in the glottal signal during voiced speech. In our process, GCI positions are used as anchor points for synchronizing residual frames. Their precise location is required, as an error on their determination might have a non-negligible impact, mainly on the subsequent PCA decomposition for the deterministic component. For this reason, SEDREAMS is used in this work as it was shown to provide high accuracy performance \cite{SEDREAMS} and directly exploits a criterion based on the discontinuity in the residual signal.

In parallel, a Mel-Generalized Cepstral (MGC) analysis is performed on the speech signals, as these features have shown their efficiency to capture the spectral envelope \cite{MGC}. As recommended in \cite{Blizzard}, we used the parameter values $\alpha=0.42$ ($Fs=16kHz$) and $\gamma=-1/3$ for MGC extraction. In this paper, we opted for the MGCs as they are widely used in speech synthesis \cite{Blizzard}, albeit other filter coefficients could be used as an alternative. Residual signals are then obtained by inverse filtering. Pitch-synchronous residual frames are finally isolated by applying a GCI-centered, two-pitch-period long Blackman windowing. The resulting dataset serves as a basis for extracting the components of the proposed DSM of the residual signal, as explained in the following sections.

\subsection{The Maximum Voiced Frequency}\label{ssec:Fm}

\begin{figure}[!ht]
  \centering
  \includegraphics[width=0.45\textwidth]{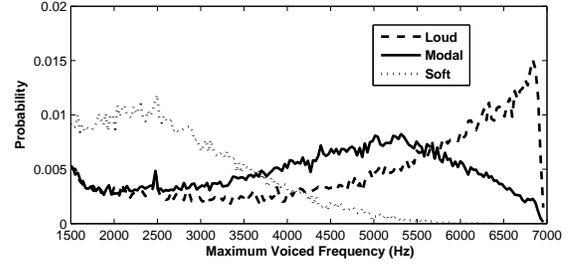}
  \caption{Histogram of the maximum voiced frequency $F_m$ for the same female speaker with three different voice qualities.}
  \label{fig:HistoFm}
\end{figure}

As previously mentioned, the DSM consists of the superposition of a deterministic $r_d(t)$ and a stochastic $r_s(t)$ component of the residual signal $r(t)$. In this model, similarly to what is done in HNM \cite{HNM}, these two contributions are supposed to hold in two distinct spectral bands. The boundary frequency between these two spectral regions is called the \emph{maximum voiced frequency} and will be denoted $F_m$ in the following. Some methods have already been proposed for estimating $F_m$ from the speech waveform \cite{HNM}, \cite{FmEstimation}. Figure \ref{fig:HistoFm} illustrates the distribution of $F_m$ estimated by the technique described in \cite{HNM} for three voice qualities (loud, modal and soft) produced by the same German female speaker. For this example, we used the De7 database originally designed for creating diphone databases for expressive speech synthesis \cite{Schroder}. A first conclusion drawn from this figure is that significant differences between the distributions are observed. More precisely, it turns out that, in general, the soft voice has a low $F_m$ (as a result of its breathy quality) and that the stronger the vocal effort, the more harmonicity in the speech signal and consequently the higher $F_m$.

However, it is worth noting that, although statistical differences are observed, obtaining a reliable trajectory of $F_m$ for a given utterance is a difficult problem \cite{YannisPhD}. For this reason, as it is done in \cite{YannisPhD} or \cite{Pantazis}, we prefer in this work to consider a fixed value of $F_m$ for a given speaker with a given voice quality (i.e assuming the voice quality constant over the considered dataset). Therefore, we use in the rest of this paper the mean value of $F_m$ extracted on a given dataset. Regarding the example of Figure \ref{fig:HistoFm}, this leads to $F_m$ = 4600 Hz for the loud, 3990 Hz for the modal and 2460 Hz for the soft voice. A disadvantage of considering a fixed value of $F_m$ is that the dynamics of the relative importance between the deterministic and stochastic components (i.e the amount of noise in speech, or its harmonicity) is not captured. On the contrary, a reliable and accurate estimation of the $F_m$ contour (which is a difficult task) is not required. Besides we observed in \cite{Drugman-Eusipco} that $F_m$ is underestimated on some speech segments, which leads after synthesis to an unpleasant predominance of noise in the speech signal. Considering a fixed value of $F_m$ alleviates this problem.

\subsection{Modeling of the Deterministic Component}\label{ssec:deterministic}

\begin{figure}[!ht]
  \centering
  \includegraphics[width=0.45\textwidth]{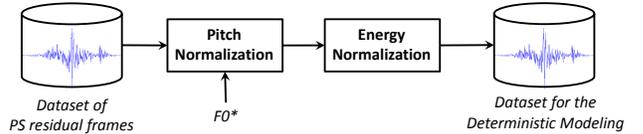}
  \caption{Workflow for obtaining the dataset for the deterministic modeling.}
  \label{fig:DetFrames}
\end{figure}

In order to model the low-frequency contents of the pitch-synchronous residual frames (extracted as explained in Section \ref{ssec:residual}), it is proposed to decompose them on an orthonormal basis obtained by Principal Component Analysis (PCA, \cite{PCA}). Preliminarily to this, the residual frames are normalized in prosody as exposed in Figure \ref{fig:DetFrames}, i.e they are normalized both in pitch period and energy. This step ensures the coherence of the dataset before applying PCA. Note that a PCA-based decomposition of glottal flow frames has been proposed in \cite{Mark-PCA}.


It is worth noticing that, for speech synthesis purpose, particular care has to be taken when choosing the number of points for length normalization. Indeed, in order to avoid the appearance of energy holes at synthesis time (occuring if the useful band of the deterministic part does not reach $F_m$ after pitch denormalization, see Section \ref{ssec:vocoder}), the pitch value $F_0^*$ for the normalization has to respect the condition:

\begin{equation}\label{eq:pitch}
F_0^* \leq \frac{F_N}{F_m} \cdot F_{0,min}
\end{equation}

where $F_N$ and $F_{0,min}$ respectively denote the Nyquist frequency and the minimum pitch value for the considered speaker. Indeed, at synthesis time, the frame normalized at $F_0^*$ and whose bandwidth reaches $F_N$ might be upsampled to $F_{0,min}$. Equation \ref{eq:pitch} guarantees that, after upsampling, its useful band still reaches $F_m$, which is the upper bound for the deterministic component. As long as $F0^*$ satisfies Equation \ref{eq:pitch}, its choice is not critical and we have verified on the Scottish male speaker AWB from the CMU ARCTIC database \cite{ARCTIC} that the effect of a change on $F0^*$ was comparable to a resampling of the resulting PCA components.

PCA can now be calculated on the resulting dataset, allowing dimensionality reduction and feature decorrelation. PCA is an orthogonal linear transformation which applies a rotation of the axis system so as to obtain the best representation of the input data, in the Least Squared (LS) sense \cite{PCA}. It can be shown that the LS criterion is equivalent to maximizing the data dispersion along the new axes. PCA can then be achieved by calculating the eigenvalues and eigenvectors of the data covariance matrix \cite{PCA}. Note that no mean removal operation is applied before PCA, which implies that obtained eigenvectors will implicitly capture the mean residual vector.

\begin{figure}[!ht]
  \centering
  \includegraphics[width=0.45\textwidth]{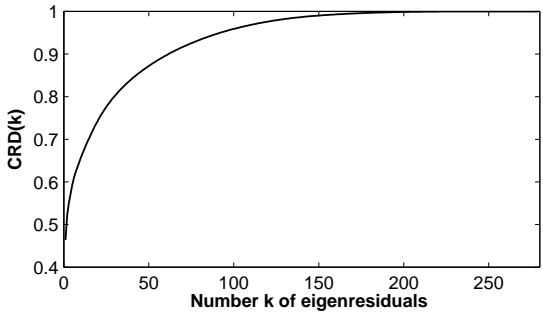}
  \caption{Evolution of the Cumulative Relative Dispersion (CRD) as a function of the number of eigenresiduals for a given male speaker.}
  \label{fig:Pval}
\end{figure}

Let us assume that the dataset consists of $N$ residual frames of $m$ samples. PCA computation will lead to $m$ eigenvalues $\lambda_i$ with their corresponding eigenvectors $\mu_i$ (here called \emph{eigenresiduals}). $\lambda_i$ is known to represent the data dispersion along axis $\mu_i$ \cite{PCA}. Using the $k$ first eigenresiduals (with $k\leq m$), the Cumulative Relative Dispersion (CRD) is defined as:

\begin{equation}\label{eq:info}
CRD(k)=\frac{\sum_{i=1}^k \lambda_i}{\sum_{i=1}^m \lambda_i},
\end{equation}

and is a relative measure of the dispersion covered over the dataset using these $k$ eigenresiduals. Figure \ref{fig:Pval} displays a typical evolution of this variable for a given male speaker ($F_s$=16kHz, $m$=280 and thus $F_0^*$=114Hz for this example). It is observed that PCA allows a high dimensionality reduction since very few eigenresiduals are sufficient to cover the greatest amount of dispersion.

\begin{figure*}[!ht]
  \centering
  \includegraphics[width=1.0\textwidth]{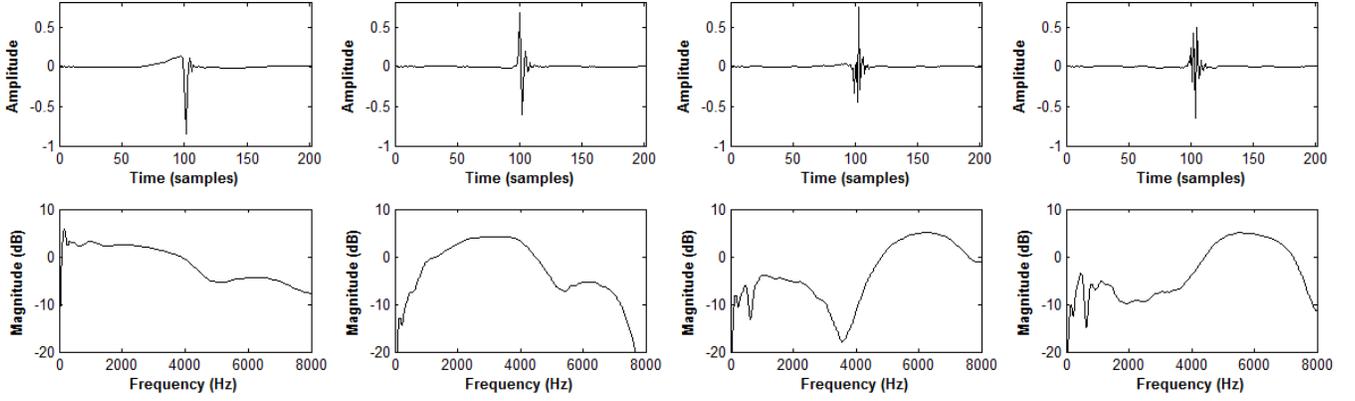}
  \caption{Illustration of the first four eigenresiduals ($\mu_1(n)$ to $\mu_4(n)$) for a given female speaker with their corresponding magnitude spectra.}
  \label{fig:4Eigenresiduals}
\end{figure*}

\begin{figure}[!ht]
  \centering
  \includegraphics[width=0.45\textwidth]{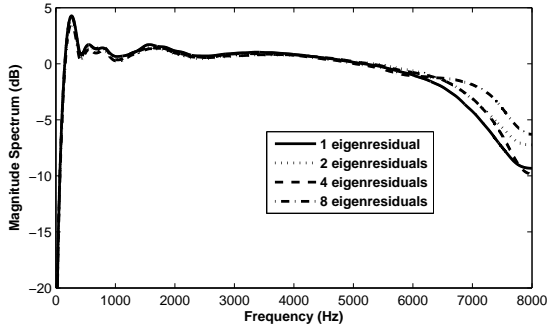}
  \caption{Reconstruction in the spectral domain of a particular residual frame (with $F0<F0^*$) using from 1 to 8 eigenresiduals.}
  \label{fig:Reconstruction}
\end{figure}

The four first eigenvectors for speaker SLT from the CMU ARCTIC database \cite{ARCTIC} are shown in Figure \ref{fig:4Eigenresiduals} together with their corresponding spectra. Note that for this example, we have chosen $F0^*$ close to $F0,min$. It is worth noting from this figure that the eigenresiduals of highest orders contribute mainly to the reconstruction of the high-frequency contents of the residual frames. In practice, we observed that, with the usual value of $F_m/F_N$, the use of only the first eigenresidual (whose relative dispersion is of 46\% in the example of Figure \ref{fig:Pval}) is sufficient for a good modeling below $F_m$, and that the effect of higher order eigenresiduals is almost negligible in that spectral band. To support this, Figure \ref{fig:Reconstruction} shows the reconstruction of a residual frame (with $F0<F0^*$) in the spectral domain using from 1 to 8 eigenvectors. As expected, it turns out that eigenresiduals of highest orders are mainly useful for reconstructing high frequencies.

Since its importance on the spectral contents below $F_m$ is predominant, the first eigenresidual $\mu_1(n)$ (just called eigenresidual for the sake of conciseness in the following) can be considered to model the deterministic component of the DSM. Besides, we will also show in the experimental parts (Sections \ref{sec:Synthesis} and \ref{sec:SpeakerReco}) that only considering the first eigenvector is sufficient for modeling the deterministic component of DSM. 




\subsection{Modeling of the Stochastic Component}\label{ssec:stochastic}

In the proposed DSM of the residual signal $r(t)$, the stochastic modeling $r_s(t)$ is similar to the noise part in the HNM \cite{HNM}. It corresponds to a white Gaussian noise $n(t)$ convolved with an auto-regressive model $h(t)$, and whose time structure is controled by an energy envelope $e(t)$:

\begin{equation}\label{eq:noise}
r_s(t)=e(t)\cdot[h(t)\star n(t)].
\end{equation}

\begin{figure}[!ht]
  \centering
  \includegraphics[width=0.45\textwidth]{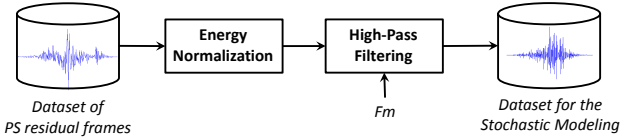}
  \caption{Workflow for obtaining the dataset for the stochastic modeling.}
  \label{fig:StochFrames}
\end{figure}

The use of $h(t)$ and $e(t)$ is required to account respectively for the spectral and temporal modulations of the high-frequency contents of the residual. In order to estimate these two contributions, the dataset of pitch-synchronous residual frames (as extracted in Section \ref{ssec:residual}) is considered, and the modifications exhibited in Figure \ref{fig:StochFrames} are brought to it. More precisely, frames are normalized in energy and only their contents beyond $F_m$ are kept. On the resulting dataset, $h(t)$ is estimated as the Linear Predictive modeling of their averaged amplitude spectrum. Indeed, since $F_m$ has been fixed and since the residual spectral envelope is almost flat over the whole frequency range, it is reasonable to consider that $h(t)$ has fairly the same effect on all frames: it acts as a high-pass filter beyond $F_m$. As for the energy envelope $e(t)$, it is determined as the average Hilbert envelope of the resulting high-filtered residual frames resampled to the normalized pitch value $F_0^*$. Note that several envelopes were studied in \cite{Pantazis} for modeling the temporal characteristics of noise in the context of HNM and for analysis-synthesis purpose. The Hilbert envelope was shown to be one of the most appropriate for this purpose.

\begin{figure*}[!ht]
  \centering
  \includegraphics[width=0.95\textwidth]{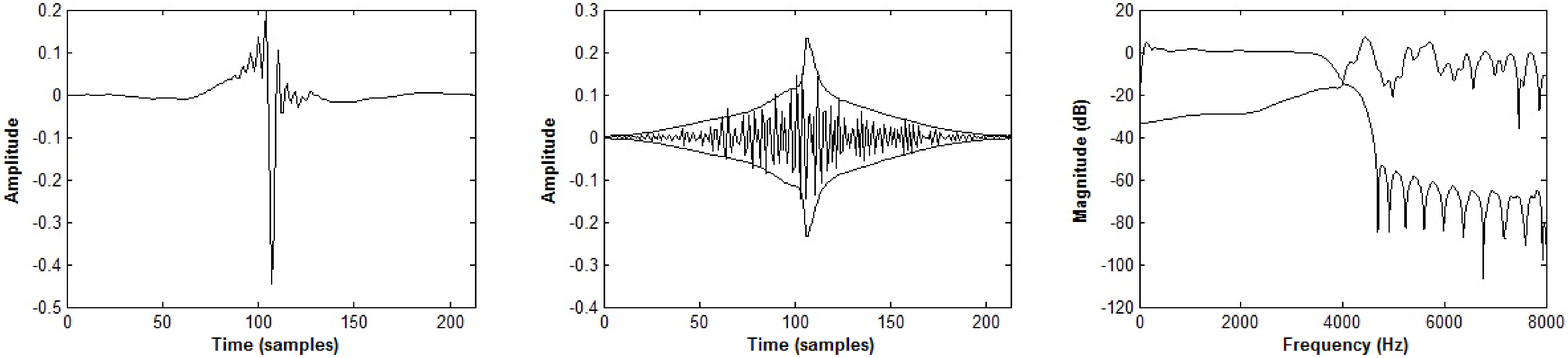}
  \caption{Illustration of the deterministic $r_d(t)$ (\emph{left plot}) and the stochastic $r_s(t)$ (\emph{mid plot}) components of DSM. The energy envelope is also indicated for information. The \emph{right plot} shows these components in the spectral domain. The deterministic part act in the low frequencies (below $F_m$) while the stochastic contribution models the contents beyond $F_m$.}
  \label{fig:DecompExample}
\end{figure*}

Figure \ref{fig:DecompExample} gives an example of DSM modeling for a particular residual frame. The two plots on the left respectively display the deterministic $r_d(t)$ and the stochastic $r_s(t)$ components constructed via the DSM vocoder (as shown in Figure \ref{fig:Vocoder}). Their contribution in the frequency domain is illustrated in the right plot of Figure \ref{fig:DecompExample}, where it is seen that both components are delimited by the maximum voiced frequency ($F_m$ = 4 kHz in this example).

\subsection{Speed of Convergence}\label{ssec:convergence}
The proposed DSM of the residual signal makes use of two important waveforms: the eigenresidual $\mu_1(n)$ for the deterministic part and the energy envelope $e(n)$ of the stochastic component. In order to estimate how much data is required for having a reliable estimation of these two signals, the male speaker AWB from the CMU ARCTIC database \cite{ARCTIC} was analyzed. This database contains about 50 minutes of speech recorded for Text-to-Speech purpose. The two reference waveforms were first computed on a large dataset containing about 150.000 pitch-synchronous residual frames. An additional estimation of these waveforms was then obtained by repeating the same operation on a held out dataset for the same speaker. The Relative Time Squared Error (RTSE) is used for both waveforms as a distance between the estimation $x_{est}(n)$ and the reference $x_{ref}(n)$ signals (where $m$ is the number of points used for pitch normalization):

\begin{equation}\label{eq:RTSE}
RTSE=\frac{\sum_{n=1}^{m}{(x_{est}(n)-x_{ref}(n))^2}}{\sum_{n=1}^{m}{x_{ref}(n)^2}}
\end{equation}

\begin{figure}[!ht]
  \centering
  \includegraphics[width=0.45\textwidth]{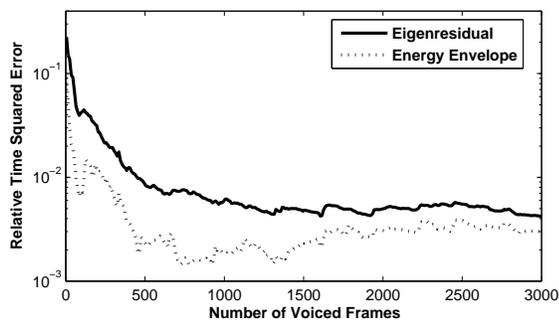}
  \caption{Speed of convergence for the eigenresidual and the energy envelope.}
  \label{fig:Convergence}    
\end{figure}

Figure \ref{fig:Convergence} displays the evolution of this measure (in logarithmic scale) with the size of the held out dataset. It may be observed that both estimations quickly converge towards the reference. From this graph, it can be considered that a dataset containing around 1000 residual frames is sufficient for obtaining a correct estimation of both the deterministic and stochastic components of the DSM. To give an idea, this corresponds to about 7s of voiced speech for a male speaker and about 4 s for a female voice.

\subsection{Phonetic Independence}\label{ssec:independence}

In the proposed DSM workflow presented in Figure \ref{fig:Vocoder}, the same modeling is used for any voiced segment. In other words, the same waveforms (eigenresidual or energy envelope) are used for the excitation of all voiced phonetic classes. In order to assess the validity of this assumption, the speaker AWB from the CMU ARCTIC database \cite{ARCTIC} was also analyzed. Reference waveforms were first extracted. In parallel, sentences were segmented into phonetic classes and for each class containing more than 1000 voiced frames (as suggested from Section \ref{ssec:convergence} for obtaining a reliable estimation), the corresponding class-dependent waveforms were calculated.


\begin{table*}[!ht]
\centering
\begin{tabular}{| c | c | c | c | c | c | c | c | c | c | c | c | c |}
\hline
\textbf{Phonetic Class} & aa & ae & ah & ao & aw & ax & ay & d & eh & er & ey & ih\\
\hline
\textbf{RTSE} & 1.16 & 0.9 & 0.8 & 1.29 & 0.81 & 0.59 & 0.84 & 9.41 & 0.81 & 0.93 & 1.18 & 0.47\\
\hline
\hline
\textbf{Phonetic Class} & iy & l & m & n & ng & ow & r & uw & v & w & y & z\\  
\hline
\textbf{RTSE} & 1.55 & 1.45 & 5.29 & 3.67 & 1.75 & 0.61 & 2.48 & 1.21 & 8.48 & 3.14 & 2.75 & 10.8\\
\hline
\end{tabular}
\caption{Relative Time Squared Error (\%) between the reference and the class-dependent first eigenresiduals.}
\label{tab:IndependenceER1}
\end{table*}

\begin{table*}[!ht]
\centering
\begin{tabular}{| c | c | c | c | c | c | c | c | c | c | c | c | c |}
\hline
\textbf{Phonetic Class} & aa & ae & ah & ao & aw & ax & ay & d & eh & er & ey & ih\\
\hline
\textbf{RTSE} & 0.32 & 0.53 & 0.32 & 0.38 & 0.38 & 0.19 & 0.4 & 5.61 & 0.77 & 0.13 & 1.23 & 0.4\\
\hline
\hline
\textbf{Phonetic Class} & iy & l & m & n & ng & ow & r & uw & v & w & y & z\\  
\hline
\textbf{RTSE} & 0.5 & 0.45 & 1.84 & 1.05 & 0.75 & 0.25 & 0.68 & 0.33 & 9.8 & 1.69 & 0.26 & 15.18\\
\hline
\end{tabular}
\caption{Relative Time Squared Error (\%) between the reference and the class-dependent energy envelopes.}
\label{tab:IndependenceEnvelope}
\end{table*}

\begin{figure}[!ht]
  \centering
  \includegraphics[width=0.5\textwidth]{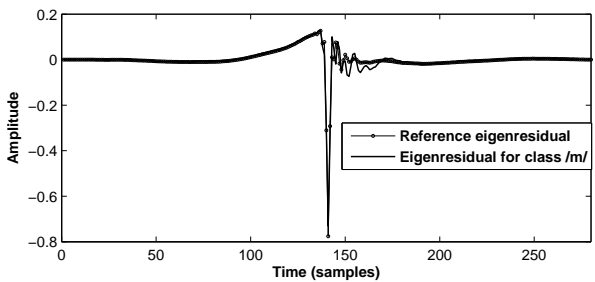}
  \caption{Waveforms of the reference eigenresidual and the one extracted on the phonetic class /m/.}
  \label{fig:ClassM}    
\end{figure}

\begin{figure}[!ht]
  \centering
  \includegraphics[width=0.5\textwidth]{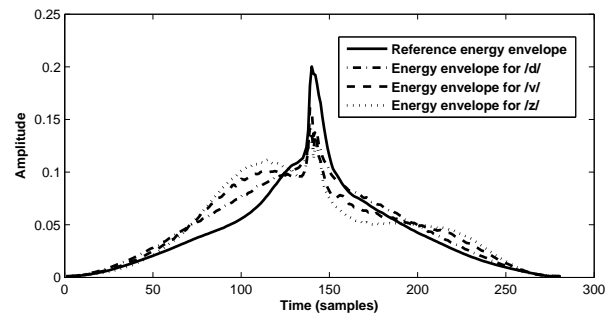}
  \caption{Waveforms of the reference energy envelope and the ones extracted on phonetic classes /d/, /v/ and /z/.}
  \label{fig:EnergyEnvelopeTot}    
\end{figure}

The values (in \%) of the RTSE between the reference and the class-dependent waveforms are shown in Tables \ref{tab:IndependenceER1} and \ref{tab:IndependenceEnvelope} respectively for the first eigenresidual and for the energy envelope. From the inspection of these results, the following conclusions can be drawn:

\begin{itemize}
\item For vowels, it can be accepted that a common modeling holds both for deterministic and stochastic components. 
\item For nasalized consonants (/m/ and /n/), RTSE reaches around 5\% on the first eigenresidual while differences are rather weak for the energy envelope. This may be explained by the difficulty in modeling the anti-formants of the vocal tract for such sounds. To illustrate the resulting differences, Figure \ref{fig:ClassM} shows the reference eigenresidual and the one extracted for the phonetic class /m/ (nasalized consonant for which the RTSE is the highest). It can be noticed that the main dissimilarities occur at the right of the GCI, while the left parts are almost identical. Indeed, according to the mixed-phase model of speech \cite{MixedPhase}, during the production of the speech signal, the response at the left of the GCI is dominated by the open phase of the glottal excitation, while the response at its right is mainly dominated by the vocal tract impulse response. After inverse filtering, the dissimilarities at the right of the GCI might then be explained by an imperfect modeling of the vocal tract transmittance for phoneme /m/.
\item For voiced fricatives and, to a lesser extent, for voiced plosives, RTSE values are not negligible for both deterministic and stochastic components and these phonetic classes could require a phone-dependent modeling in the DSM framework. Figure \ref{fig:EnergyEnvelopeTot} displays the reference energy envelope and the ones extracted for the phonetic classes /d/, /v/ and /z/ (for which RTSEs are the highests). It is observed that the noise distribution is significantly altered for these phonemes. Among others, it can be noticed that the energy envelope for /d/ is more spread than the reference, while for /v/ and /z/ there is a clear asymmetry with higher energy on the left of the GCI (i.e around the glottal open phase).

\end{itemize}

Nonetheless, in the rest of this manuscript, a phone-independent approach will be adopted. In other words, a single waveform for the first eigenresidual and for the energy envelope will be used for all phonetic classes. Phonetic dependence is indeed not an issue for speaker recognition (Section \ref{sec:SpeakerReco}), in which phonemes cannot assume to be known. For speech synthesis purpose (Section \ref{sec:Synthesis}), our attempts to integrate a phone-dependent modeling in the DSM vocoder did not bring any audible differences through our informal tests. Therefore the assumption of using a common modeling independent of the phonetic context is supposed to hold in the rest of this paper. Note also that these conclusions have been drawn from our analysis for the English and French languages. In a general way, there may be other languages for which this conclusion might not hold. For example, this might be the case in Finnish due to the presence of vocal fry.

\section{Application of DSM to Speech Synthesis}\label{sec:Synthesis}

This Section discusses how the proposed DSM of the residual signal can be useful in parametric speech synthesis. First, the principle of the DSM vocoder is presented in Section \ref{ssec:vocoder}. Details about our HMM-based speech synthesizer using the DSM excitation model are given in Section \ref{ssec:HTSintegration}. Finally Section \ref{ssec:HTSexp} gives, in the context of HMM-based speech synthesis, a subjective evaluation between the proposed DSM and two other methods of excitation modeling: the traditional pulse source and the STRAIGHT technique.

\subsection{The DSM Vocoder}\label{ssec:vocoder}

A workflow summarizing the proposed DSM vocoder has been presented in Figure \ref{fig:Vocoder}. The vocoder takes only two feature streams as input: pitch values ($F_0$) for the source, and MGC coefficients for the filter (with $\alpha=0.42$ and $\gamma=-1/3$, as indicated in Section \ref{ssec:residual}). All other data ($F_m$, $F_0^*$, the first eigenresidual, the energy envelope and the autoregressive model for the stochastic component) is precomputed on a training dataset as explained in Section \ref{sec:DSM}. As our informal attempts showed that adding eigenresiduals of higher orders has almost no audible effect on the delivered speech synthesis, only the first eigenresidual is considered for speech synthesis purpose. This was also the case in our preliminary tests \cite{DSM} where using the 15 first eigenresiduals did not lead to a significant modification in the reconstruction of the deterministic component of the residual frames and provided sensibly a similar synthesis quality.

The deterministic component $r_d(t)$ of the residual signal then consists of the (first) eigenresidual resampled such that its length is twice the target pitch period. Following Equation \ref{eq:noise}, the stochastic part $r_s(t)$ is a white noise modulated by the autogressive model and multiplied in time by the energy envelope centered on the current GCI. Note that the energy envelope is also resampled to the target pitch. Both components are then overlap-added so as to obtain the residual signal $r(t)$. In the case of unvoiced regions, the excitation merely consists of white Gaussian noise. The synthesized excitation is finally the input of the Mel-Log Spectrum Approximation (MLSA, \cite{MLSA}) filter to generate the final speech signal.

\subsection{HMM-based Speech Synthesis based on DSM}\label{ssec:HTSintegration}

\begin{figure}[!ht]
  \centering
  \includegraphics[width=0.3\textwidth]{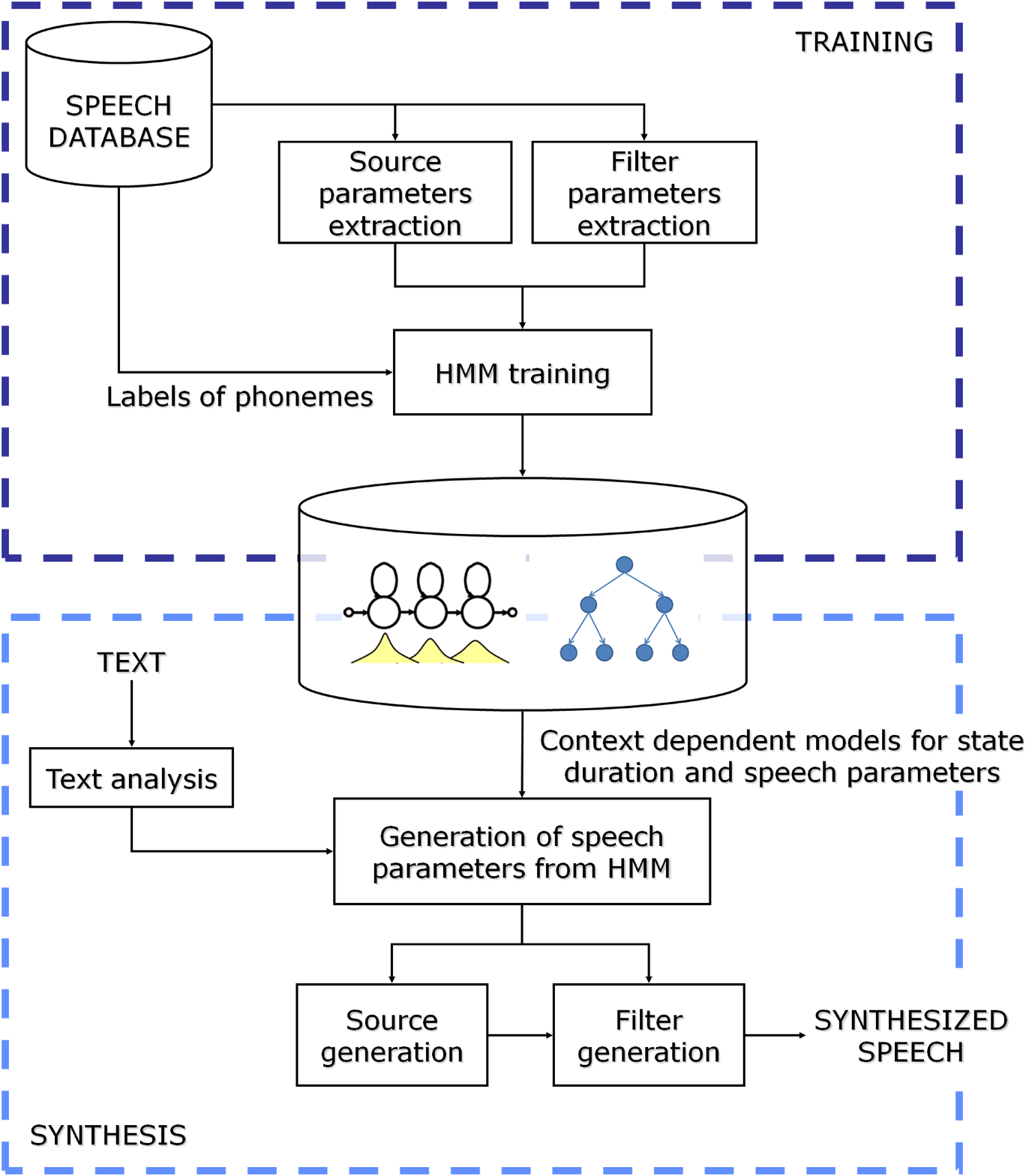}
  \caption{Framework of a HMM-based speech synthesizer (adapted from \cite{SPS}).}
  \label{fig:HTS}    
\end{figure}

HMM-based speech synthesis aims at generating natural sequences of speech parameters directly from a statistical model, which is previously trained on a given speech database \cite{SPS}. The general framework of a HMM-based speech synthesizer is displayed in Figure \ref{fig:HTS}. Two main steps can be distinguished in this process: training and synthesis.

The \textbf{training} step assumes that a large segmented speech database is available. Labels consist of a phonetic environment description. First, both excitation (source) and spectral (filter) parameters are extracted from the speech signals. Since source modeling may be composed of either continuous values or a discrete symbol (respectively during voiced and unvoiced regions), Multi-Space probability Density (MSD) HMMs have been proposed \cite{MSD}, as this approach is able to model sequences of observations having a variable dimensionality. Given the speech parameters and the labels, HMMs are trained using the Viterbi and Baum-Welch re-estimation algorithms \cite{SPS}. Decision tree-based context clustering is used to statistically model data appearing in similar contextual situations. Indeed contextual factors such as stress-related, locational, syntaxical or phonetic factors affect prosodic (duration and source excitation characteristics) as well as spectral features. More precisely an exhaustive list of possible contextual questions is first drawn up. Decision trees are then built for source, spectrum and duration independently using a maximum likelihood criterion. Probability densities for each tree leaf are finally approximated by a Gaussian mixture model.


At \textbf{synthesis} time, the input text is converted into a sequence of contextual labels using a Natural Language Processor. From them, a path through the context-dependent HMMs is computed using the duration decision tree. Excitation and spectral parameters are then generated by maximizing the output probability. The incorporation of dynamic features ($\Delta$ and $\Delta^2$) makes the coefficients evolution more realistic and smooth \cite{GenHMM}. The generated parameters are then the input of the vocoder, which produces the synthetic waveform.

The implementation of our HMM-based speech synthesizer relies on the HTS toolkit publicly available in \cite{HTS}. As mentioned in Section \ref{ssec:vocoder}, the only excitation feature used for the training is $F_0$. A five-state left-to-right multistream HMM is used. More precisely, four separate streams are employed: \emph{i)} one single Gaussian distribution with diagonal covariance for the spectral coefficients and their derivatives, \emph{ii)} one MSD distribution for pitch, \emph{iii)} one MSD distribution for pitch first derivative, and \emph{iv)} one MSD distribution for pitch second derivative. In each MSD distribution, for voiced parts, parameters are modeled by single Gaussian distributions with diagonal covariance, while the voiced/unvoiced decision is modeled by an MSD weight. As HMMs are known for oversmoothing the generated trajectories \cite{GV}, the Global Variance technique \cite{GV} is used to alleviate this effect. The generated parameters are then fed into the vocoder described in Section \ref{ssec:vocoder}.

\subsection{Experiments of HMM-based speech synthesis}\label{ssec:HTSexp}

In this Section, the proposed DSM is compared to two other well-known excitation models for HMM-based speech synthesis purpose. The first method is the traditional \emph{Pulse} excitation, used by default in the HTS toolkit \cite{HTS}. This technique basically uses either a pulse train during voiced speech, or white noise during unvoiced parts. The resulting excitation signal is then the input of the MLSA filter.

\begin{figure}[!ht]
  \centering
  \includegraphics[width=0.45\textwidth]{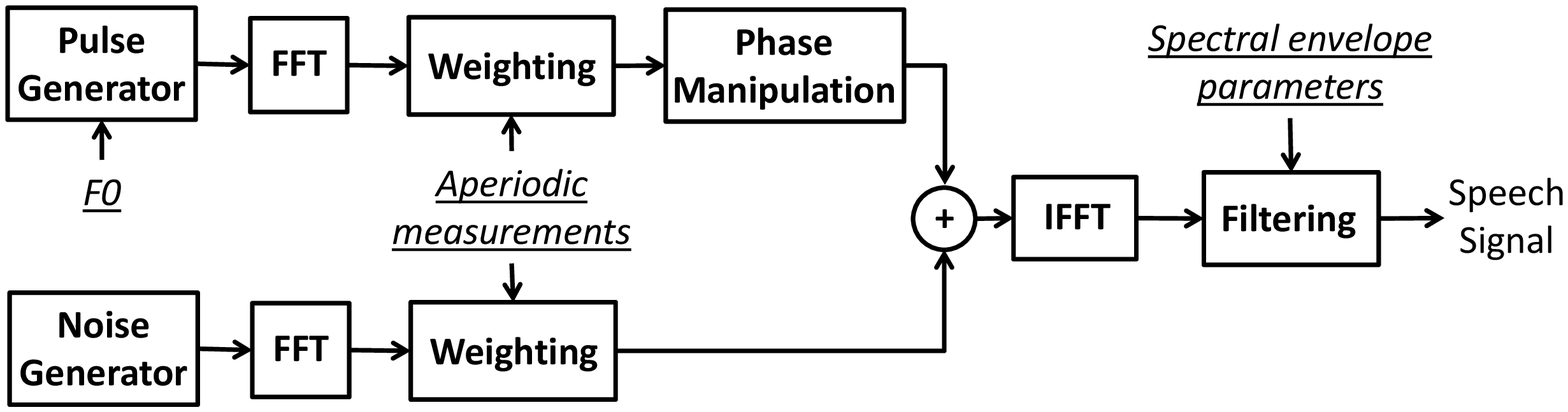}
  \caption{Workflow of the STRAIGHT vocoder. Input features (indicated in italic and underlined) are, for the excitation, the target pitch $F_0$ and the 5 aperiodic measurements, and the spectral envelope parameters for the filter.}
  \label{fig:STRAIGHTVocoder}    
\end{figure}

The second method is the \emph{STRAIGHT} vocoder, known for its high-quality representation of the speech signal. STRAIGHT makes use of a specific spectral envelope obtained via a pitch-adaptive time-frequency smoothing of the FFT speech spectrum. As for the excitation modeling, STRAIGHT relies on aperiodic measurements in five spectral subbands: [0-1], [1-2], [2-4], [4-6] and [6-8] kHz. As a consequence, the excitation features used by the HMM synthesizer now include the 5 aperiodic measurements besides $F_0$. This results in an additional HMM stream composed of these aperiodicity parameters, together with their first and second derivatives. Once generated, the speech features are the input of the STRAIGHT vocoder presented in Figure \ref{fig:STRAIGHTVocoder}. The source signal is a Mixed Excitation whose periodic and aperiodic components are weighted by the aperiodicity measures. As suggested in \cite{STRAIGHT}, the phase of the periodic contribution is manipulated so as to reduce buzziness. Both components are then added and passed through a minimum-phase filter obtained from the parameters describing the smooth STRAIGHT spectral envelope.

\subsubsection{Experimental Protocol}\label{sssec:HTSprotocol}

The synthetic voices of two UK English speakers were assessed. The first is a male speaker who recorded about ten hours, while the second is a female speaker with about four hours of speech. The HMM-based speech synthesizers were trained, for both voices, on the whole corpus. This was carried out, as explained in Section \ref{ssec:HTSintegration}, by the Centre for Speech Technology Research of Edinburgh, which kindly provided us the generated parameters. All details about the training and the parameter generation can be found in \cite{JoaoThesis}, as well as other experiments with other excitation models. 

The test consists of a subjective comparison between the proposed DSM and both the traditional pulse excitation and STRAIGHT. More precisely, we performed a Comparative Mean Opinion Score (CMOS,\cite{CMOS}) test composed of 23 sentences, among which the first 3 were provided for calibration. These utterances were randomly chosen out of a set of 120 sentences, half for each speaker. For each sentence, participants were asked to listen to both versions (DSM versus Pulse or STRAIGHT, randomly shuffled) and to attribute a score according to their overall preference. The CMOS scale is a 7-point scale ranging from -3 (DSM is much worse than the other method) to +3 (meaning the opposite). A null score is given if the quality of both versions is found to be equivalent. A positive score means that DSM is preferred over the other technique, a negative one implying the opposite. Participants were divided into two categories: 26 speech experts (i.e people familiar with speech processing) and 34 naive listeners. The test was conducted through the Web.

\subsubsection{Results}\label{sssec:HTSresults}

Results of the CMOS test are exhibited in Table \ref{tab:CMOSResults} and are separated for the two categories of participants. First, it is observed that speech experts significantly perferred DSM over the pulse excitation, with a CMOS score of a bit more than 1.2 for both the male and the female speaker. A similar conclusion can be drawn for the naive listeners, although their averaged CMOS scores are around 0.75 instead of 1.2. As a matter of fact, we observed that naive listeners used the whole CMOS scale in a lesser extent. Indeed, since the only change between the two versions only concerns the excitation modeling (as spectral envelope and prosody were kept unchanged), auditive differences were relatively subtle. It can then be understood that speech experts noticed them more easily. Regarding the comparison with STRAIGHT, it turns out that both methods were found, in average, to deliver a comparable quality. Although speech experts very slightly preferred DSM, the opposite is noted for naive listeners. But taking the 95\% confidence intervals into account, no significant advantage for DSM over STRAIGHT, or vice versa, can be highlighted.

\begin{table}[!ht]
\centering
\begin{tabular}{ c | c | c |}
\textbf{Speech Experts}  & Male Speaker & Female Speaker \\
\hline
DSM vs Pulse & \textbf{1.205} $\pm$ 0.198 & \textbf{1.241} $\pm$ 0.209 \\
\hline
DSM vs STRAIGHT & \textbf{0.167} $\pm$ 0.217 & \textbf{0.037} $\pm$ 0.197 \\
\hline
\hline
\textbf{Naive listeners}  & Male Speaker & Female Speaker \\
\hline
DSM vs Pulse & \textbf{0.75} $\pm$ 0.176 & \textbf{0.722} $\pm$ 0.188 \\
\hline
DSM vs STRAIGHT & \textbf{-0.010} $\pm$ 0.164 & \textbf{-0.072} $\pm$ 0.201 \\ 
\hline
\end{tabular}
\caption{Average CMOS scores together with their 95 \% confidence intervals, for both speech experts and naive listeners.}
\label{tab:CMOSResults}
\end{table}

\begin{table*}[!ht]
\centering
\begin{tabular}{ c || c | c | c || c | c | c |}
  & \multicolumn{3}{|c||}{Male Speaker} & \multicolumn{3}{|c|}{Female Speaker}\\
\hline
 \textbf{Speech Experts} & DSM preferred & Equivalent & Other method preferred & DSM preferred & Equivalent & Other method preferred \\ 
\hline
DSM vs Pulse & 76.07 \% & 18.80 \% & 5.13 \% & 73.68 \% & 18.80 \% & 7.52 \% \\
\hline
DSM vs STRAIGHT & 33.33 \% & 40.35 \% & 26.32 \% & 35.77 \% & 32.85 \% & 31.39 \% \\
\hline
\hline
 \textbf{Naive listeners} & DSM preferred & Equivalent & Other method preferred & DSM preferred & Equivalent & Other method preferred \\ 
\hline
DSM vs Pulse & 59.38 \% & 24.38 \% & 16.24 \% & 62.78 \% & 16.11 \% & 21.11 \% \\
\hline
DSM vs STRAIGHT & 31.22 \% & 33.86 \% & 34.92 \% & 30.47 \% & 33.11 \% & 36.42 \% \\
\hline
\end{tabular}
\caption{Preference scores for both speech experts and naive listeners.}
\label{tab:PrefResults}
\end{table*}

In complement to the CMOS scores, Table \ref{tab:PrefResults} presents the preference results for all test conditions. While speech experts preferred DSM to Pulse in about 75\% of cases, this proportion is reduced to around 60\% for naive listeners. Nevertheless, the advantage of DSM over Pulse is clear again, as Pulse was only preferred in very few cases. Regarding the comparison with STRAIGHT, preference results confirm that both methods are almost equivalent. Indeed it is seen in Table \ref{tab:PrefResults} that the repartition between the three categories is almost one third, reflecting the fact that both methods lead to a similar quality. However, one advantage of DSM over STRAIGHT is that it does not require the addition of a specific stream in the HMM-based synthesizer, making not only the training step lighter, but also more importantly alleviating the computational footprint at running time. On the other hand, DSM relies on several non-parametric characteristics, which makes it poorly flexible.

In \cite{DSM}, we also carried out a subjective evaluation between DSM and Pulse, for 5 English and French voices. The CMOS test was submitted to 40 people, among them both speech experts and naive listeners. Since the data, the synthesizer itself and the test conditions are not the same, results are obviously not directly comparable. However, the conclusions drawn from these two experiments both report the overwhelming advantage of DSM over Pulse in speech synthesis. This superiority was even stronger in \cite{DSM} where the averaged CMOS scores varied between 1 and 1.8 across the 5 voices and the preference rates for DSM between 78\% and 94\%.

Finally, note that we performed in \cite{Drugman-Eusipco} a comparative evaluation of several techniques for pitch modification, as a preliminary step for voice transformation \emph{in an analysis-synthesis context}. DSM was, among others, compared to HNM and STRAIGHT. It turned out from that study that, in terms of overall quality, DSM outperformed HNM for both male and female speakers, and STRAIGHT for male voices, while STRAIGHT gave the best performance on female speakers.

\section{Application of DSM to Speaker Recognition}\label{sec:SpeakerReco}

Automatic speaker recognition refers to the use of a machine in order to recognize a person from a spoken phrase \cite{Reynolds}. This task is then closely linked to the understanding of what defines the speaker individuality. Although high-level information (such as the word usage) could be of interest, low-level acoustic features are generally employed \cite{Reynolds}. Such features are most of the time extracted from the amplitude spectrum of the speech signal. They aim at parameterizing the contribution of the vocal tract, which is an important characteristic of the speaker identity.

On the other hand, very few works address the possibility of using features derived from the glottal source in speaker recognition. In \cite{Thevenaz}, Thevenaz exploits the orthogonality of the LPC residue for text-independent speaker verification. In order to avoid synchronization with pitch epochs and simultaneously to get rid of the residual phase contribution, it was suggested to retain the residual amplitude spectrum. It is concluded in that paper that although the residue-based features are less informative than the vocal tract-based ones, they are nonetheless useful for speaker verification, and above all combine favourably with methods based on the LPC filter. The approach proposed in \cite{Prasanna} extracts speaker-specific information from several consecutive cycles (typically 60) of the residual signal using auto-associative neural networks. In \cite{Murty}, Murty et al. demonstrate the complementarity of features based on the residual phase with the traditional MFCCs, commonly used in speaker recognition. Authors led speaker recognition experiments on the NIST-2003 database. By integrating the residual phase information in addition to the common MFCCs, they reported a reduction of equal error rate from 14\% to 10.5\%. In \cite{Plumpe}, Plumpe et al. focused on the use of the glottal flow estimated by closed phase inverse filtering. On the resulting glottal source, two types of features were extracted. The first ones are time-domain features, parameterizing both the coarse structure (obtained by fitting a LF model \cite{LF}) and the fine structure of the glottal flow derivative. The second ones are a Mel-cepstral representation of the glottal source. A clear advantage in favor of the cepstral coefficients was shown. In a similar way, Gudnason et al. focus in \cite{Gudnason} on the use of Voice Source Cepstrum Coefficients (VSCCs) for speaker recognition. A process based on closed-phase inverse filtering, and which is shown to be robust to LPC analysis errors and low-frequency phase distortion, is proposed. When combined to traditional MFCCs, the resulting features are reported to lead to an appreciable improvement for speaker identification.

The goal of this section is to investigate the usefulness of the proposed DSM of the residual excitation for speaker recognition purpose. For this, we suggest to use the speaker-dependent waveforms of the DSM, as introduced in Section \ref{sec:DSM}: the eigenresiduals for the deterministic part and the energy envelope for the stochastic contribution. These waveforms are also called \emph{glottal signatures} in the following, as they are glottal-based signals conveying a relevant amount of information about the speaker identity. Note that the whole identification process described in the following is fully automatic, i.e no manual correction is applied on the GCI positions, the determination of $F_0$ or the voicing contours, as explained in Section \ref{sec:DSM}.

This section is structured as follows. Section \ref{ssec:Incorporation} explains how the DSM-based waveforms are used for speaker identification purpose. In Section \ref{ssec:ExpSpeakerReco}, the protocol used for our experiments is described. Section \ref{ssec:TIMIT} presents our results on the large TIMIT database. First of all, the potential of the proposed waveforms is investigated, as well as the impact of the higher orders eigenresiduals. Then, speaker identification performance using the glottal signatures is assessed. Our experiments on the 
 database are reported in Section \ref{ssec:YOHO}. This gives an idea of the inter-session sensitivity of the proposed technique. On both databases, some comparisons with other glottal-based speaker recognition approaches \cite{Plumpe}, \cite{Gudnason} are provided.

\subsection{Integrating Glottal Signatures in Speaker Identification}\label{ssec:Incorporation}

In order to be integrated into a speaker identification system, the proposed DSM-based signatures are estimated on both training and testing sets.  A \emph{distance matrix} $D(i,j)$ between speaker $i$ (whose glottal signatures are estimated on the training dataset) and speaker $j$ (estimated on the testing dataset) is then computed. In this work, the RTSE (see Equation \ref{eq:RTSE}) is chosen as a distance measure between two waveforms. Finally, the identification of a speaker $i$ is carried out by looking for the lowest value in the $i^{th}$ row of the distance matrix $D(i,j)$. The speaker is then correctly identified if the position of the minimum is $i$. In other words, when a new recording is presented to the system, the identified speaker is the one whose glottal signatures are the closest (in the Euclidian sense) to the signatures extracted on this recording.


In the following, it will be observed that no more than two glottal signatures are used for speaker identification. Many strategies are possible for combining their information and draw a final decision \cite{Kittler}. In this study, two strategies are considered: a weighted multiplication or a weighted sum. More precisely, denoting $D_x(i,j)$ and $D_y(i,j)$ the distance matrices using respectively the glottal signatures $x(n)$ and $y(n)$, the two sources of information are merged in our framework by calculating the final distance matrix $D(i,j)$ respectively as:

\begin{equation}\label{eq:Multiplication}
D(i,j)=D_x(i,j)^{\alpha} \cdot D_y(i,j)^{1-\alpha}
\end{equation}

\begin{equation}\label{eq:Addition}
D(i,j)=\beta \cdot D_x(i,j) + (1-\beta) \cdot D_y(i,j)
\end{equation}

where $\alpha$ and $\beta$ are weights ranging from 0 to 1. They are used to possibly emphasize the importance of a given glottal signature with regard to the other. When the weight is 0, only $y(n)$ is considered, while a weight equal to 1 means that only $x(n)$ is used for identification.

\subsection{Experimental Protocol}\label{ssec:ExpSpeakerReco}

In this Section, the maximum voiced frequency $F_m$ is fixed to 4 kHz (usual value for a modal voice quality, as shown in Section \ref{ssec:Fm}) and the normalized pitch value $F_0^*$ is set to 100 Hz for all speakers. Albeit these two parameters are known to be speaker-dependent (as explained in Section \ref{sec:DSM}), fixing a common value across all speakers is required in order to match the glottal signatures. Experiments are carried out on both TIMIT and YOHO databases, for comparison purpose with \cite{Plumpe} and \cite{Gudnason}. In \cite{Plumpe}, Plumpe et al. reported speaker identification results on TIMIT using either Time-Domain features (TDGF) or a Mel-Cepstral (MCGF) representation of the estimated Glottal Flow. As for \cite{Gudnason}, Gudnason et al. performed tests on both TIMIT and YOHO using their proposed Voice Source Cepstrum Coefficients (VSCC). For both methods, classification was performed using a GMM-based approach.

The TIMIT database \cite{TIMIT} comprises 10 recordings from 630 speakers (438 males, 192 females) sampled at 16 kHz, with about 30 seconds of speech per speaker. As for the YOHO database \cite{YOHO}, it contains speech from 138 speakers (108 males, 30 females) sampled at 8 kHz. Since $F_s = 8 kHz$ for YOHO, only the deterministic part of the DSM holds, and the unvoiced energy envelope cannot therefore be used for the recognition. Recordings of YOHO were collected in a real-world office environment through 4 sessions over a 3 month period. For each session, 24 phrases were uttered by each speaker.

In the following experiments, the data is split for each speaker (and each session for YOHO) into 2 equal parts for training and testing. This is done in order to guarantee that, for both steps, enough residual frames are available for reliably estimating the signatures (see Section \ref{ssec:convergence}). However, it is worth noting that although there was always a sufficient number of frames for YOHO, it happened for some low-pitched voices of the TIMIT database that the amount of available data was rather limited (typically only around 500 voiced frames were used for these voices for the training or the test). This consequently led to an imperfect estimation of the glottal signatures in such cases.

\subsection{Results on the TIMIT database}\label{ssec:TIMIT}

\subsubsection{Usefulness of the glottal signatures}\label{sssec:Potential}

\begin{figure}[!ht]
  \centering
  \includegraphics[width=0.45\textwidth]{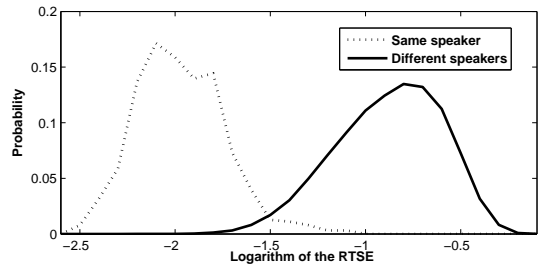}
  \caption{Distributions of the Relative Time Squared Error (RTSE) between the first eigenresiduals $\mu_1(n)$ estimated respectively for the same speaker and for different speakers.}
  \label{fig:ErrorDistrib}    
\end{figure}

To give a first idea on the potential of using the glottal signatures in speaker recognition, Figure \ref{fig:ErrorDistrib} displays the distributions of $D_{\mu_1}(i,j)$ (i.e the distance matrix using only the first eigenresidual $\mu_1(n)$) respectively when $i=j$ and when $i\ne j$. In other words, this plot shows the histograms of the RTSE (in logarithmic scale) between the first eigenresiduals estimated respectively for the same speaker and for different speakers. It is clearly observed that the error measure is much higher (about 15 times higher in average) when the tested signature does not belong to the considered speaker. It is also noticed that, for the same speaker, the RTSE on the eigenresidual is about $1\%$, which is of the same order of magnitude as for the inherent estimation process, confirming our results of Sections \ref{ssec:convergence} and \ref{ssec:independence}. However a weak overlap between both distributions is noted, which may lead to some errors in terms of speaker identification.

\subsubsection{Effect of the higher order eigenresiduals}\label{sssec:HigherOrders}

\begin{figure}[!ht]
  \centering
  \includegraphics[width=0.45\textwidth]{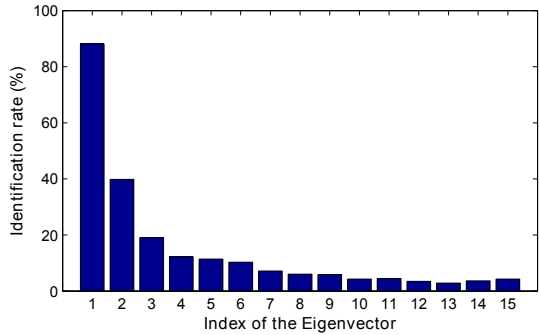}
  \caption{Speaker identification capability on the whole TIMIT database using \emph{individually} the eigenresiduals of higher orders.}
  \label{fig:Reco_Eigenvectors}    
\end{figure}

It was mentioned in Section \ref{ssec:deterministic}, that only considering the first eigenresidual is sufficient for a good modeling of the residual signal below $F_m$, and that the effect of higher order eigenresiduals is almost negligible in that spectral band. One could argue however that higher order waveforms can be useful for speaker recognition. Figure \ref{fig:Reco_Eigenvectors} shows the identification rate on the whole TIMIT database (630 speakers), for each eigenresidual $\mu_i(n)$. It is clearly observed that higher order eigenresiduals are less discriminative about the speaker identity. More particularly, the identification rate dramatically drops from 88.6\% to 39.8\% when going from the first to the second eigenresidual used individually. 

\begin{figure}[!ht]
  \centering
  \includegraphics[width=0.5\textwidth]{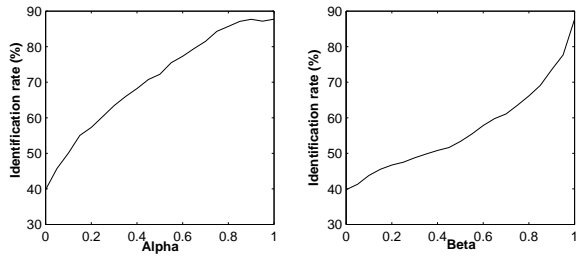}
  \caption{Evolution of the identification rate as a function of $\alpha$ and $\beta$, when \textbf{the first and second eigenresiduals ($\mu_1(n)$ and $\mu_2(n)$) are combined} according to Equations (\ref{eq:Multiplication}) and (\ref{eq:Addition}).}
  \label{fig:CombineER1andER2}    
\end{figure}

In order to assess the contribution of higher order eigenresiduals, Figure \ref{fig:CombineER1andER2} shows the evolution of the identification rate as a function of $\alpha$ and $\beta$, when the first and second eigenresiduals $\mu_1(n)$ and $\mu_2(n)$ are combined according to Equations (\ref{eq:Multiplication}) and (\ref{eq:Addition}). In both strategies, it turns out that considering $\mu_2(n)$ in addition to $\mu_1(n)$ does not bring anything, since optimal performance is reached for $\alpha$=1 and $\beta$=1. Therefore, the effect of higher order eigenresiduals for speaker identification can be neglected and only $\mu_1(n)$ is considered in the following experiments.

\subsubsection{Combining the eigenresidual and the energy envelope}\label{sssec:ER1andE}

\begin{figure}[!ht]
  \centering
  \includegraphics[width=0.5\textwidth]{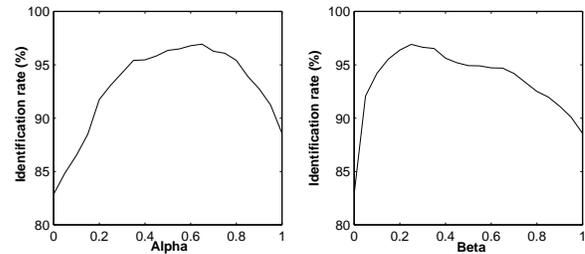}
  \caption{Evolution of the identification rate as a function of $\alpha$ and $\beta$, when \textbf{the first eigenresidual $\mu_1(n)$ and the energy envelope $e(n)$ are combined} according to Equations (\ref{eq:Multiplication}) and (\ref{eq:Addition}).}
  \label{fig:CombineER1andEnvelope}    
\end{figure}

Contrarily to higher order eigenresiduals, the energy envelope $e(n)$ of the stochastic part (see Section \ref{ssec:stochastic}) showed a high discrimination power with an identification rate of 82.86\% on the whole TIMIT database. It can then be expected that using the first eigenresidual $\mu_1(n)$ in complement to $e(n)$ could improve the performance. For this, they are combined as in Equations (\ref{eq:Multiplication}) and (\ref{eq:Addition}), and the influence of $\alpha$ and $\beta$ is displayed in Figure \ref{fig:CombineER1andEnvelope}. First, the advantage of using both signatures together is clearly confirmed. Secondly, the optimal performance using Eq. (\ref{eq:Multiplication}) or Eq. (\ref{eq:Addition}) is identical. In the rest of our experiments, we used Equation \ref{eq:Multiplication} with $\alpha$=0.5 which, although slightly suboptimal in this example, makes the combination as a simple element-by-element multiplication of $D_{\mu_1}(i,j)$ and $D_e(i,j)$.

\subsubsection{Speaker identification results}\label{sssec:IDresults}

\begin{figure}[!ht]
  \centering
  \includegraphics[width=0.45\textwidth]{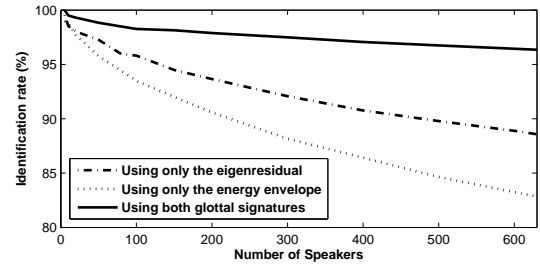}
  \caption{Evolution of the identification rate with the number of speakers for the TIMIT database.}
  \label{fig:SpeakerInfluence}    
\end{figure}

Figure \ref{fig:SpeakerInfluence} exhibits the evolution of the identification rate with the number of speakers considered in the database. Identification was achieved using only one of the two glottal signatures, or using their combination as suggested in Section \ref{ssec:Incorporation}. As expected the performance drops as the number of speakers increases, since the risk of confusion becomes more important. However this degradation is relatively slow in all cases. One other important observation is the clear advantage of combining the information of the two signatures. Indeed this leads to an improvement of $7.78\%$ compared to using only the first eigenresidual on the whole database.  

Table \ref{tab:TIMITresults} summarizes the results obtained on the TIMIT database. Identification rates for 168 speakers are also given for comparison purpose. Using the time-domain parametrization of the glottal flow (TDGF), Plumpe et al. \cite{Plumpe} reported an average misclassification rate of $28.65\%$. This result was importantly reduced to $4.70\%$ by making use of the Mel-cepstral representation of the glottal flow (MCGF). On the same subset, Gudnason et al. reported in \cite{Gudnason} a misclassification rate of $5.06\%$ using their proposed VSCC. These results can be compared to the $1.98\%$ we achieved using the two glottal signatures. Finally also note that, relying on the VSCC,  Gudnason et al. \cite{Gudnason} obtained a misidentification rate of $12.95\%$ on the whole TIMIT database (630 speakers). With the proposed signatures, a misclassification rate of $3.65\%$ is reached. It is worth noting that no specific disparity bewteen male and female speakers was observed. More precisely, 6 out of the 192 female speakers ($3.13\%$), and 17 out of the 438 male speakers ($3.88\%$) were misclassified using the two glottal signatures.

\begin{table}[!ht]
\centering
\begin{tabular}{c | c | c}
 & 168 speakers & 630 speakers \\
\hline
\hline
TDGF \cite{Plumpe} & 28.65 & / \\
\hline
MCGF \cite{Plumpe} & 4.70 & / \\
\hline
VSCC \cite{Gudnason} & 5.06 & 12.95 \\
\hline
Using only the eigenresidual & 5.88 & 11.43 \\
\hline
Using only the energy envelope & 8.76 & 17.14\\
\hline
Using both glottal signatures & \textbf{1.98} & \textbf{3.65} \\
\hline
\end{tabular}
\caption{Misidentification rate ($\%$) on the TIMIT database obtained using state-of-the-art glottal approaches or the proposed DSM-based signatures.}
\label{tab:TIMITresults}
\end{table}

\subsection{Results on the YOHO database}\label{ssec:YOHO}

As mentioned above, recordings in the YOHO database are sampled at 8 kHz, and therefore only the first eigenresidual is used for speaker identification. Besides, as the 4 sessions were spaced over 3 months, we evaluate here the inter-session variability of the proposed glottal signature. Table \ref{tab:YOHOresults} reports our speaker identification results as a function of the period separating the training and testing sessions. In addition the proportions of cases for which the correct speaker is recognized in second or third position (instead of first position) are also given. When recordings are from the same session, an almost perfect performance is carried out, with 99.73\% of correct identification. This is above the approximative 95\% rate reached on TIMIT with the eigenresidual for the same number of speakers (see Figure \ref{fig:SpeakerInfluence}). This might be explained by the greater amount of data available in YOHO for the estimation of the glottal signature.

On the contrary, when the test is performed one session later, the identification dramatically drops by $30\%$. This first degradation accounts for two phenomena: the mismatch between training and testing recording conditions, and the intra-speaker variability. It then turns out that the identification rate decreases of about $5\%$ for any later session. This is mainly attributable to speaker variability, which increases with the period separating the two sessions. As future work, we plan to design some channel compensation in order to alleviate the mismatch between training and testing sessions. Indeed different recording conditions impose different characteristics to the speech signal. Among these, differences in phase response may dramatically affect the estimation of the signatures (since the information of the residual is essentially contained in its phase).

It is worth noting that when recording sessions differ, between $13\%$ and $16\%$ of speakers are identified in second or third position. By integrating a complementary source of information, such as the traditional features describing the vocal tract function, it can be expected that most of the ambiguity on these signatures will be removed. Finally note that Gudnason et al. reported in \cite{Gudnason} an identification rate of $63.7\%$ using the VSCC, but with test recordings coming from the 4 sessions. By averaging our results over all sessions, the use of only the eigenresidual leads to an identification rate of $71.1\%$, confirming the good performance of the DSM-based signatures for speaker recognition.

\begin{table}[!ht]
\centering
\begin{tabular}{c || c | c | c}
 & First Position & Second Position & Third Position \\
\hline
\hline
Same session & 99.73\% & 0.27\% & 0\% \\
\hline
One session later & 69.29\% & 7.88\% & 5.19\% \\
\hline
Two sessions later & 64.31\% & 8.83\% & 4.57\% \\
\hline
Three sessions later & 58.70\% & 11.78\% & 4.35\% \\
\hline
\end{tabular}
\caption{Proportion of speakers classified in first (correct identification), second and third position, when recordings are spaced over several sessions.}
\label{tab:YOHOresults}
\end{table}

\section{Conclusion}\label{sec:conclu}
This paper presented a new excitation model: the Deterministic plus Stochastic Model (DSM) of the residual signal. DSM estimation is performed by automatic analysis of a speaker-dependent dataset of pitch-synchronous residual frames. After a detailed description of the underlying theoretical framework, some computational and phonetic considerations were examined. It was proved that a speaker-dependent dataset of around 1000 voiced frames is sufficient for having a reliable estimation of the DSM components. It was also shown that the assumption of considering a common excitation modeling for all phonetic classes is valid. The applicability of the proposed DSM was then studied for two major fields of speech processing: speech synthesis and speaker recognition.


First, the DSM vocoder was integrated into a HMM-based speech synthesizer. The quality delivered by the resulting synthesizer was compared to HMM-based synthesis using either the traditional pulse excitation or the STRAIGHT method. A subjective comparative evaluation on two speakers (male and female) was performed by 60 listeners, among them speech experts and naive listeners. Results showed a significant preference for DSM over the pulse excitation, this advantage being clearer for speech experts. Regarding the comparison with STRAIGHT, both techniques turned out to lead to similar quality.

Secondly, the usefulness of glottal signatures derived from the proposed DSM for speaker identification was investigated. Their potential use was studied on the large TIMIT database, and the recognition carried out relying on DSM-based signatures was observed to outperform by large the use of other glottal-based features proposed in the literature. Finally, in a second test on the YOHO database, we evaluated the inter-session sensitivity of these signatures, highlighting the degradation due to a mismatch between recording conditions, and the intra-speaker variability.

As a result from the experiments in both speech synthesis and speaker identification, it can be concluded that two glottal signatures are sufficient within the proposed DSM: the first eigenresidual for the deterministic part, and the energy envelope for the stochastic part.

There are several possible extensions to the current version of DSM. The dynamics of the maximum voiced frequency could be taken into account. In our system, filter parameters are estimated asynchronously. Nonetheless, we could benefit from the knowledge of GCI positions to perform a pitch-synchronous analysis which could improve the modeling of the spectral envelope. 

As future works, we plan to investigate speaker recognition strategies for integrating the proposed DSM-based glottal signatures within a traditional approach relying on features related to the spectral envelope. This problem is more complex than it may seem since our proposed system is based on a matching error, while traditional techniques are based on a probabilistic GMM approach. However, this approach seems promising as these two sources of information are complementary. In addition, automatic phase compensation could alleviate the mismatch between recording conditions. 

\section*{Acknowledgment}

Thomas Drugman is supported by the Belgian Fonds National de la Recherche Scientifique (FNRS). We would like to thank Dr. Cabral and Prof. Renals at the Centre for Speech Technology Research of Edinburgh for their precious help on the speech synthesis experiments. We are also grateful to Geoffrey Wilfart and Acapela Group for helping us in the first synthesis tests. Authors also would like to thank reviewers for their fruitful comments and suggestions.


%



\ifCLASSOPTIONcaptionsoff
  \newpage
\fi



%
\bibliographystyle{IEEEtran}
\bibliography{bare_jrnl}

\end{document}